\documentclass[twocolumn,floatfix,nofootinbib,aps]{revtex4}
\usepackage{graphicx}
\usepackage{subfig}
\usepackage{bm}
\usepackage{ulem}
\usepackage{dcolumn}
\usepackage{hyperref}
\usepackage{cancel}
\usepackage{color}
\usepackage{latexsym}
\usepackage{nicefrac}
\usepackage{latexsym}
\usepackage{amsmath, mathrsfs, bm}

\begin{document}

\title{Constraining the interaction between dark sectors with future HI intensity mapping observations}

\author{Xiaodong Xu$^{1}$, Yin-Zhe Ma$^{2,3}$, and Amanda Weltman$^{1,4}$}
\affiliation{$^{1}$Department of Mathematics and Applied Mathematics, University of Cape Town, Rondebosch 7701, Cape Town, South Africa}
\affiliation{$^{2}$School of Chemistry and Physics, University of KwaZulu-Natal, Durban 4000, South Africa}
\affiliation{$^{3}$NAOC-UKZN Computational Astrophysics Centre (NUCAC), University of KwaZulu-Natal, Durban, 4000 South Africa}
\affiliation{$^{4}$ School of Natural Sciences, Institute for Advanced Study, Olden Lane, Princeton, New Jersey 08540, USA}


\begin{abstract}
We study a model of interacting dark matter and dark energy, in which the two components are coupled. We calculate the predictions for the 21-cm intensity mapping power spectra, and forecast the detectability with future single-dish intensity mapping surveys (BINGO, FAST and SKA-I). Since dark energy is turned on at $z\sim 1$, which falls into the sensitivity range of these radio surveys, the HI intensity mapping technique is an efficient tool to constrain the interaction. By comparing with current constraints on dark sector interactions, we find that future radio surveys will produce tight and reliable constraints on the coupling parameters.
\end{abstract}

\maketitle

\section{Introduction \label{sec.intro}}

One of the central problems in modern cosmology is the mystery of the observed accelerated expansion of the Universe. After almost two decades of this discovery, the simplest and most successful candidate is still the cosmological constant, i.e. ``$\Lambda$'' term in Einstein's field equation. The $\Lambda$CDM model, in which the late time dynamics of the Universe is described by a cosmological constant together with cold dark matter (DM), is generally consistent with all current observations~\cite{PlanckCollaboration2015}, although recently there is some hint that dark energy might be dynamical~\cite{Zhao2017}. In spite of its observational success, the $\Lambda$CDM model is challenged with some theoretical difficulties: the mechanism with which the cosmological constant can emerge from fundamental physics with the observed value is lacking \cite{Weinberg1989, Martin2012} and without dynamics, the coincidence problem remains.

To solve these problems, a plethora of alternative proposals have been introduced in the literature. In a large class of the proposals, dynamical dark energy (DE), which is an exotic form of matter whose equation of state (EoS) is close to $-1$ at late times, is responsible for the accelerated expansion. There are a bunch of fully studied dynamical DE models, including quintessence \cite{Caldwell1997}, K-essence \cite{Armendariz-Picon2001}, chameleon fields \cite{Khoury2003} and Horndeski theory \cite{Deffayet2011, Horndeski1974}, many of which describe DE with a scalar field. While most dynamic DE models ignore the so-called phantom regime, where the DE EoS $w<-1$, some specific theory such as braneworld and Brans-Dicke gravity can lead to phantom DE \cite{Sahni2003, Elizalde2004}. Apart from the dynamics of DE, it is also interesting to consider the coupling of the DE with DM, as an interaction between DE and DM. If such a coupling exists, it may affect the dynamics of the Universe given their dominance in the matter budget today.

Both DE and DM have not been detected directly, and the study of the interaction between them is based on astronomical observations. It is then difficult to describe them, especially DE, from first principles. Dark energy is often treated as a perfect fluid, and its interaction with DM is conveniently parameterized with a phenomenological model. The presence of interactions within the dark sector inevitably modifies the evolution of the Universe, in both expansion history and formation of large scale structures. At the background level, the phenomenological interacting dark energy (IDE) models can have tracker solutions both in the past and in the future ~\cite{Zimdahl2001, Chimento2003, Amendola2006a, Wei2006, DelCampo2006, Guo2007a, Feng2007, Feng2008}. It is, however, possible for a model with a varying EoS to reproduce the same expansion history as in the IDE model. The evolution of density perturbations, large scale structure formation and signatures in observations in the IDE scenario are investigated using perturbation theory \cite{He2008a, He2009a, He2009, He2010a, Xu2011}. In several cases, the IDE models have been found to have perturbations with unstable growing modes. The stability of perturbations depends on both the parametrization of the interaction and dark energy EoS \cite{Valiviita2008, He2008a, Corasaniti2008, Jackson2009, Xu2011}. Readers can refer to Ref.~\cite{Wang2016} for a recent review on theoretical challenges, cosmological implications and observational signatures in IDE models.

By comparing the predictions of the IDE model with observations we can constrain the interaction term. The most precise and information-rich cosmological observations to date come from the cosmic microwave background (CMB) experiments~\cite{PlanckCollaboration2015a}. However, most of the information imprinted in the CMB map dated to the last scattering surface, where the redshift $z \sim 1100$, while DE was always subdominant until the late time accelerated expansion era, where $z \lesssim 1$. Thus they are only moderately efficient in terms of constraining the interaction between dark sectors. Using  low redshift observations, including the measurements of the local Hubble parameter, type-Ia supernova data (SNIa), baryon acoustic oscillations (BAO), growth rate of large scale structure, weak lensing etc. can also set constraints on the IDE models, despite the fact that current late time observations are not comparable with CMB experiments in providing high precision constraints. The interaction between dark sectors was tested against CMB measurements, low redshift observations and their combination and found consistent with current observational data~\cite{Costa2017, Marttens2017, Yang2017, Marcondes2016, Murgia2016, Pu2015, Li2015, Yang2015, Salvatelli2014, Li2013, Xu2013, Xia2013, Salvatelli2013}. Some recent observational data indicate that a moderate interaction may be present at late times \cite{Salvatelli2014}. In addition, several researches reported that the presence of the interaction can alleviate the tension on the value of the Hubble constant $H_0$ between the CMB anisotropy constraints obtained from the {\it Planck} satellite and the recent direct measurements \cite{DiValentino2017, PlanckCollaboration2015, Riess2016}.

If we are only interested in measuring the matter distribution on large scales, it is not necessary to resolve individual galaxies. The 21-cm intensity mapping (IM) technique is to conduct a survey with relatively low angular resolution that detects only the integrated intensity from many unresolved sources. This technique allows for extremely large volumes to be surveyed very efficiently~\cite{Bigot-Sazy2015a}. For example, the SKA will be capable of performing large IM surveys over $0 \lesssim z \lesssim 3$ by using the redshifted neutral hydrogen (HI) 21-cm emission line. FAST will be able to create a large area HI map at redshift $0.05 \lesssim z \lesssim 0.35$~\cite{Bigot-Sazy2015}. In the late Universe, neutral hydrogen resides mainly in dense regions inside galaxies that are shielded from the ionizing UV background, and the 21-cm line is narrow and relatively unaffected by absorption or contamination by other lines. The 21-cm radiation is thus an excellent tracer of the matter density field in redshift space. For a detailed discussion on the uses of line-intensity mapping, see \cite{Kovetz2017}. The ongoing radio surveys promise to provide 21-cm IM data that contain information of the same amount or even more than current CMB observations at late times \cite{CameraSantosFerreiraEtAl2013, Xu2014a, Bull2015}.

HI IM is a natural tool in studying dark energy phenomenon. Comparing to other probes, such as galaxy surveys, it has several advantages. It can probe deeper redshift than typical galaxy surveys with thin redshift bins, leading to good observations of the evolution of both background expansion and the growth of large structure formation. Unlike galaxy surveys, HI IM does not measure an individual galaxy's redshift and position. Rather, it measures the total flux of HI emission on scales of BAO, and surveys a large volume. So there are more signals coming from large, linear scales where the effect of dark energy is the most important. In this work, we examine the ability of this promising tool, which seemingly provides the most information from the redshift and scales that DE is sensitive to, in constraining the interaction between DM and DE.

This paper is organized as follows. In Secs.~\ref{sec.IDE} and \ref{sec.power21cm}, we briefly review the IDE model and compute the 21-cm angular power spectrum. In~Sec.~\ref{sec.survey}, we analyse the experimental noise and survey parameters for the three single-dish 21-cm IM telescopes, BINGO, FAST and SKA-I. We then forecast the constraints on the interaction by using a Fisher matrix analysis in Sec.~\ref{sec.forecast}. Our conclusion is presented in Sec.~\ref{sec.conclusion}.

Besides the interaction parameters, we assume the other cosmological parameters to be {\it Planck} 2015 best-fitting values~\cite{PlanckCollaboration2015} if not specified otherwise: $h=0.6774$, $\Omega_{\mathrm{m}}h^2=0.1188$, $\Omega_{\mathrm{b}}h^{2}=0.0223$, $\ln({10^{10}A_{\mathrm{s}}})=3.064$ and $n_{\mathrm{s}}=0.9667$.

\section{The interacting dark energy model \label{sec.IDE}}

We consider a cosmological model with a phenomenologically inspired interaction between DM and DE \cite{He2010a}. In this scenario, the energy momentum tensor of DM and DE is not conserved separately, but satisfies the following equation,
\begin{equation}
\nabla_{\mu} T_{\lambda}^{\mu\nu} = Q_{\lambda}^{\nu},
\label{eq.EM_conservation}
\end{equation}
where the subscript $\lambda$ refers to either DM($\mathrm{c}$) or DE($\mathrm{d}$). $Q_{(\lambda)}^{\nu}$ is the coupling vector representing the interaction between DM and DE. We assume the dark sector does not interact nongravitationally with normal matter; thus the total energy momentum of the dark sectors is conserved, so that $Q_{\mathrm{c}}^{\nu} + Q_{\mathrm{d}}^{\nu} = 0$.

We assume that the Universe is described by a Friedmann-Lema\^{i}tre-Robertson-Walker metric with small perturbations over a smooth background. The line element is given by
\begin{eqnarray}
\mathrm{d}s^2 &=& a^2 [ -(1+2\psi)\mathrm{d}\tau^2 + 2\partial_iB\mathrm{d}\tau\mathrm{d}x^i \nonumber \\ &+& (1+2\phi)\delta_{ij}\mathrm{d}x^i\mathrm{d}x^j
+ (\partial_i\partial_j - \frac{1}{3}\delta_{ij}\nabla^2)E\mathrm{d}x^i\mathrm{d}x^j ], \nonumber \\
\end{eqnarray}
where $\tau$ refers to the conformal time. Functions $\psi$, $B$, $\phi$ and $E$ are the perturbations to the metric which are not all independent of each other. Therefore we could have different gauge choices by selecting different functions in the perturbed metric. In the following, we work in Newtonian gauge where $ B=E=0$~\cite{Costa2013}. The choice of a particular gauge does not affect the predictions on observables such as the angular power spectrum in linear perturbation regime \cite{Kodama:1985bj}.

The matter components in the Universe are described by the energy momentum tensor of a perfect fluid
\begin{equation}
T^{\mu\nu} = (\rho+p) U^{\mu}U^{\nu} + pg^{\mu\nu}.
\end{equation}

Due to a lack of understandings of either DM or DE at present, it is difficult to postulate their coupling from first principles. Hence we instead describe the interaction phenomenologically and focus on its impact on the dynamics of the Universe rather than the microscopic mechanism. In this work, we assume that the interaction appears as energy transfer between DM and DE which is proportional to their energy densities in the homogeneous and isotropic background. The equations of motion of DM/DE energy densities read
\begin{eqnarray}
&& \dot{\rho_{\mathrm{c}}} + 3\frac{\dot{a}}{a}\rho_{\mathrm{c}} = a^2Q_{\mathrm{c}}^0 = aQ_{\mathrm{c}}, \label{eq.rhoc}\\
&& \dot{\rho_{\mathrm{d}}} + 3\frac{\dot{a}}{a}(1+w)\rho_{\mathrm{d}} = a^2Q_{\mathrm{d}}^0 = aQ_{\mathrm{d}}, \label{eq.rhod}
\end{eqnarray}
where
\begin{equation}
Q_{\mathrm{c,d}} \equiv g_{\mu\nu} Q_{\mathrm{c,d}}^{\mu} Q_{\mathrm{c,d}}^{\nu}.
\end{equation}
In our model,
\begin{equation}
Q_{\mathrm{c}} = -Q_{\mathrm{d}} = 3H(\xi_1\rho_{\mathrm{c}} + \xi_2\rho_{\mathrm{d}}).
\label{eq.Q}
\end{equation}
$w$ is the EoS of DE and $\xi_1$ and $\xi_2$ are dimensionless coupling coefficients. Due to its phenomenological nature, the IDE model can accommodate both quintessence ($w>-1$) and phantom ($w<-1$) DE. A dot denotes a derivative with respect to conformal time. Here we assume the coupling coefficients are constant for simplicity. This allows us to compare our forecast for HI IM observations with various constraints with existing data in the literatures, which adopt the same assumption. We compare such constraints in Sec.~\ref{sec.forecast}. The linear perturbation to the zeroth component of the coupling vector can be derived from Eq.~(\ref{eq.Q}),
\begin{equation}
\delta Q_{(\lambda)}^0 = -\frac{\psi}{a}Q_{(\lambda)} + \frac{1}{a}\delta Q_{(\lambda)},
\end{equation}
while the $i$th component needs to be specified in the model in addition to the background energy transfer. We assume $\delta Q_{(\lambda)}^i$ vanishes, which implies that there is no scattering between DM and DE; only an inertial drag effect appears due to stationary energy transfer. While it is a simple parametrization of a DM-DE interaction, which assumes the energy transfer is proportional to the energy density of DM and/or DE, we can put some constraints on the coupling constants as well as DE EoS $w$ by some physical considerations. First, to avoid the unphysical solution of a negative DE density ($\rho_{\mathrm{d}} < 0$) in the early Universe, the coupling constant $\xi_1$ must be positive \cite{He2008}. Furthermore, it has been found that, when $\xi_1 \neq 0$, the curvature perturbations diverge in early times if $w+1$ is a positive constant \cite{Xu2011}. Thus we exclude cases (a) $w> -1$, $\xi_1 \neq 0$ and (b) $w<-1$, $\xi_1<0$ in our study.

\section{The power spectrum of 21-cm radiation \label{sec.power21cm}}

The 21-cm line corresponds to the transition between the fundamental hyperfine levels of neutral hydrogen atoms, which corresponds to the frequency $\nu = 1420$ MHz in the rest frame. The brightness temperature fluctuation of the redshifted 21-cm signal reflects the distributions of HI, which serves as a tracer of the three-dimensional large scale structure of the Universe. The observed brightness temperature at redshift $z$ is given by~\cite{HallBonvinChallinor2012}
\begin{equation}
T_{\mathrm{b}}(z,\hat{\mathbf{n}}) = \frac{3}{32\pi} \frac{(h_{\mathrm{p}}c)^3n_{\mathrm{HI}}A_{10}}{k_{\mathrm{B}}E_{21}} \bigg\vert\frac{\mathrm{d}\lambda}{\mathrm{d}z}\bigg\vert,
\end{equation}
where $h_{\mathrm{p}}$ is Planck's constant; $k_{\rm B}$ is Boltzman's constant; $A_{10}=2.869\times 10^{15}\,{\rm s}^{-1}$ is the spontaneous emission coefficient; $E_{21}=5.88\,\mu{\rm eV}$ is the rest frame energy of the 21-cm transition; $n_{\mathrm{HI}}$ is the number density of neutral hydrogen atoms at a given redshift; $\hat{\mathbf{n}}$ is the unit vector along the line of sight and $\lambda$ is an affine parameter of the propagation of photons. Ignoring the perturbations, in the homogeneous and isotropic background, the brightness temperature can be written as
\begin{eqnarray}
\bar{T}_{\mathrm{b}}(z) &=& \frac{3}{32\pi} \frac{(h_{\mathrm{p}}c)^3\bar{n}_{\mathrm{HI}}A_{10}}{k_{\mathrm{B}}E_{21}^2(1+z)H(z)} \\
    &=& 0.188h \Omega_{\mathrm{HI}}(z) \frac{(1+z)^2}{E(z)} \mathrm{K},
\end{eqnarray}
where $E(z) \equiv H(z) / H_0$, $H_0 = 100 h\,\mathrm{km~s}^{-1}\mathrm{Mpc}^{-1}$ is the Hubble parameter at present, and $\Omega_{\mathrm{HI}}$ is the fractional density of neutral hydrogen in the Universe. $\Omega_{\mathrm{HI}}$ can safely be assumed to be constant in the redshift range ($z \leq 3$) considered. Besides, the 21-cm power spectrum is also modulated by HI bias $b_{\mathrm{HI}}$. In this work, we assume $b_{\mathrm{HI}}$ is a constant independent of scales. we take $\Omega_{\mathrm{HI}} = 0.62 \times 10^{-3}$~\cite{Prochaska2008,Switzer2013} and $b_{\mathrm{HI}}=1$ throughout this work if not stated otherwise.

Perturbing $T_{\mathrm{b}}$ to linear order, the angular cross power spectrum between two redshift windows is~\cite{Li2017}
\begin{equation}
C_{\ell}^{ij} = 4\pi \int\mathrm{d}\ln{k} P_{\mathcal{R}}(k) \Delta^i_{T_{\mathrm{b}},\ell}(k) \Delta^j_{T_{\mathrm{b}},\ell}(k).
\label{eq.Cl_th}
\end{equation}
$P_{\mathcal{R}}(k)$ is the power spectrum of the dimensionless primordial curvature fluctuation $\mathcal{R}$, and $\Delta^i_{T_{\mathrm{b}},\ell}(k) = \Delta^i_{T_{\mathrm{b}},\ell}(\mathbf{k}) / \mathcal{R}(\mathbf{k})$. The sources are integrated over a redshift window function $W(z)$
\begin{equation}
\Delta^i_{T_{\mathrm{b}},\ell}(\mathbf{k}) = \int\mathrm{d}z W_{i}(z) \Delta_{T_{\mathrm{b}},\ell}(\mathbf{k},z)
\end{equation}
$\Delta_{T_{\mathrm{b}},\ell}(\mathbf{k},z)$ is the spherical harmonic expansion of fluctuations to the brightness temperature. We use a Gaussian window function in the following calculations.

\begin{figure*}[tp]
\subfloat[]{
    \includegraphics[width=0.45\textwidth]{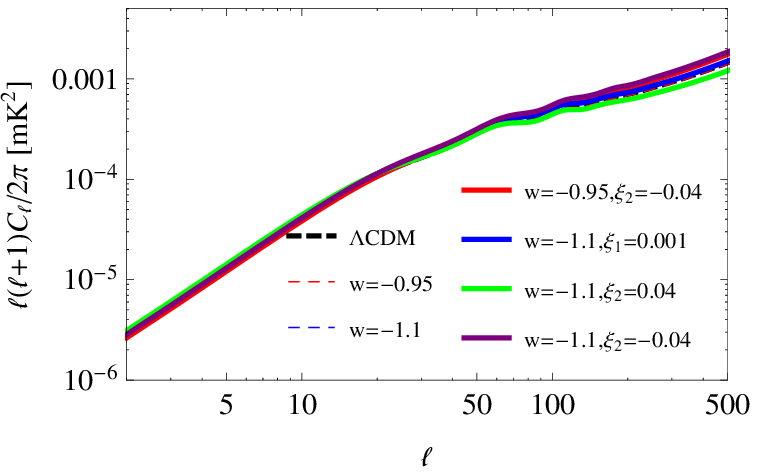}
}
\subfloat[]{
    \includegraphics[width=0.43\textwidth]{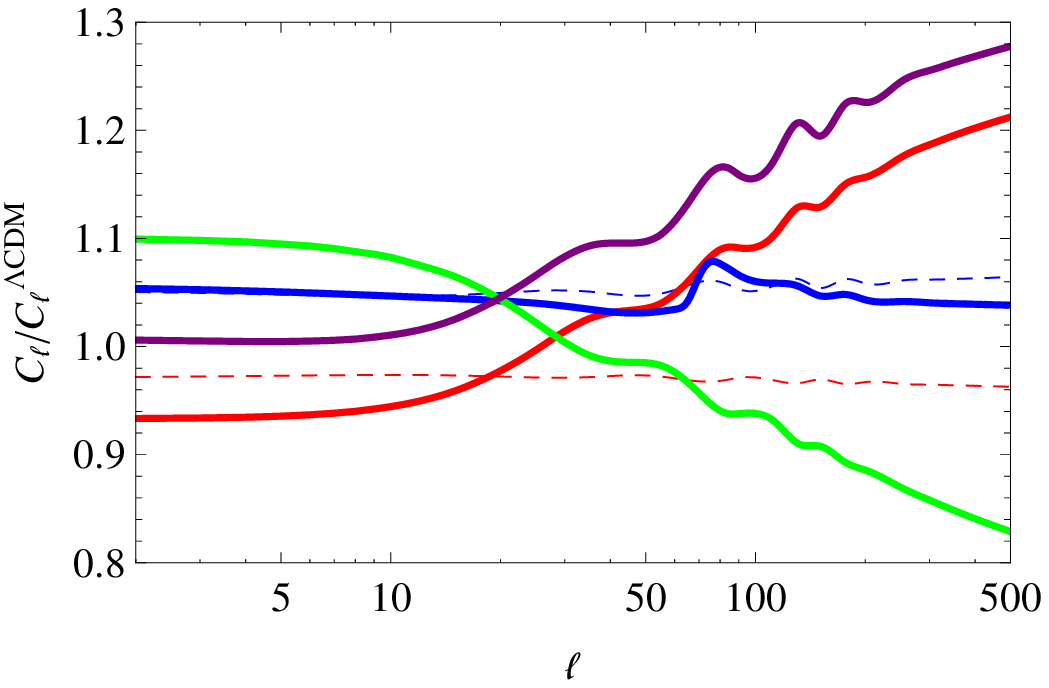}
}
\caption{\label{fig.cl} (a) The auto power spectra of 21-cm in $\Lambda$CDM, $w$DE (varying $w$ but fixed $\xi_{1}=\xi_{2}=0$) and IDE models (varying $w$, $\xi_{1}$ and $\xi_{2}$) for a $\Delta\nu = 20\mathrm{MHz}$ window centered at $z=0.3$ ($\nu=1092\,$MHz). (b) The ratio of $C_{\ell}$'s in $w$DE and IDE models with respect to the $\Lambda$CDM model. The legend is shown in panel (a). }
\end{figure*}

Figure~\ref{fig.cl} displays the autocorrelation angular power spectra at $z=3$ for the fiducial $\Lambda$CDM, dynamical DE model with constant EoS ($w$DE) and IDE models. The coupling strength $\xi_1$ and $\xi_2$ in the IDE models are tuned to a similar magnitude as the best-fitting values of the {\it Planck} 2015 data \cite{Costa2017}. The deviations of the angular power spectra in IDE models from $\Lambda$CDM demonstrate strong scale dependence and can be up to 20\% at small angular scales. Furthermore, the wiggles in $C_{\ell}$ around $\ell \sim 100$ indicate the shift of the BAO scale in the presence of a dark interaction.

\section{Surveys \label{sec.survey}}

Given that the interaction within the dark sector significantly changes the 21-cm brightness temperature angular power spectrum, the HI IM method would serve as an efficient probe of this interaction. In order to forecast the future detectability of this interaction, here we investigate the following future surveys of HI IM from three different radio telescopes.

\begin{itemize}
\item BAO from Integrated Neutral Gas Observations (BINGO) is a single-dish IM project \cite{Battye2016, Battye2013}. BINGO will work on the frequency from 960 to 1260 MHz, making an overall instantaneous bandwidth of 300 MHz. The working frequency corresponds to a redshift range of $z \simeq 0.13$--$0.48$. The telescope will comprise two static dishes, in which one acts as a secondary -- the so-called crossed-dragone/compact range antenna design -- and perform a drift scan. The 56 receivers arranged at the focal plane will create an instantaneous field of view of about $10^{\circ}$(in declination direction)$\times 9^{\circ}$(in right ascension direction), pointing at $\sim -45^{\circ}$ declination.

\item The Five-hundred-meter Aperture Spherical Radio Telescope (FAST) is the largest single-dish radio telescope in the world with an active primary surface of 300 meter  diameter, which is expected to achieve the highest sensitivity within its frequency bands of any single-dish radio telescopes \cite{Li2016, Nan2011}. FAST will be able to slew between sources in less than 10 minutes, while the maximum slew is $80^{\circ}$. Hence, FAST will work in a drift survey mode similar to BINGO, but adjust the zenith angle range from Dec:$-14^{\circ}12'$ to Dec:$65^{\circ}48'$, covering about 24000 deg$^2$ sky area in one year operation. FAST will be equipped with receivers covering the frequency range from 70 MHz to 3 GHz, among which the 19 beam feed horn array at L-band will be the primary survey instrument. We consider the frequency range 1050--1350 MHz in our analysis, corresponding to $z \simeq 0.05$--$0.35$.

\item The Square Kilometre Array (SKA) project will be the world's largest radio telescope, with eventually over a square kilometer of collecting area.\footnote{http://www.skatelescope.org} For SKA Phase 1 (SKA-I), the site in South Africa's Karoo desert will host about 200 mid to high frequency instruments, each having a movable dish of 15 m in diameter. The SKA midfrequency antennae will cover a wide frequency range from 350 MHz ($z \simeq 3.0$) upwards and the survey will cover a sky area of about 25000 deg$^2$. In our analysis, we consider only the single-dish mode survey and stacked data of each dish together, for consistency with the other projects. However, the angular resolution in single-dish mode is significantly lower than the interferometry mode, which reduces the sensitivity at large $\ell$'s.
\end{itemize}

\begin{table*}[tb!]
\caption{\label{tab.survey_parameter} Survey parameters.}
\begin{tabular}{p{140pt}ccc}
    \toprule
    & BINGO & FAST & SKA-I \\
    \hline
    Frequency range (MHz)                        & [960, 1260] & [1050, 1350] & [350, 1050] \\
    Redshift range                               & [0.13, 0.48]& [0.05, 0.35] & [0.35, 3.06] \\
    System temperature $T_{\mathrm{sys}}$ (K)    & 50          & 25           & 28 \\
    Number of dishes $N_{\mathrm{d}}$            & 1           & 1            & 190 \\
    Number of beams $N_{\mathrm{b}}$             & 56          & 19           & 1 \\
    Illuminated aperture $D_{\mathrm{dish}}$ (m) & 25          & 300          & 15 \\
    Sky coverage $A_{\mathrm{sky}}$ (deg$^2$)    & 3000        & 24000        & 25000 \\
    Observation time $t_{\mathrm{obs}}$ (yr)     & 1           & 1            & 1 \\
    Frequency window width (MHz)                 & 20          & 20           & 20 \\
    \lasthline
\end{tabular}
\end{table*}

The experimental parameters employed in our analysis are shown in Table~\ref{tab.survey_parameter}. The observed frequencies in 21-cm radiation projects are much lower than that in CMB experiments, and the problem of contamination due to foregrounds, such as galactic synchrotron emission, extragalactic point sources, and atmospheric signal, is much more severe in HI IM. It is then necessary to employ foreground removal techniques to reduce the foreground contamination \cite{Bigot-Sazy2015a, Olivari2015, Zhang2015}. In reality, there is always some residual foreground even after applying such techniques to the maps. Yet, we assume perfect foreground removal and consider only instrumental noise in the analysis. In this optimistic case, the cross correlations of the noise between different frequency (redshift) windows are assumed to be negligible. The instrumental noise is dominated by the thermal noise which can be calculated as~\cite{Bull2015}
\begin{equation}
N_{\ell} = \frac{T_{\mathrm{sys}}^2 A_{\mathrm{sky}}}{N_{\mathrm{d}} N_{\mathrm{b}} t_{\mathrm{obs}} \Delta\nu},
\label{eq.noise}
\end{equation}
where $T_{\rm sys}$ is the system temperature, $A_{\rm sky}$ is the survey area, $N_{\rm d}$ and $N_{\rm b}$ are the number of dishes and number of antennae in each dish respectively, $t_{\rm obs}$ is the total observational time (integration time) and $\Delta \nu$ is the frequency window (frequency resolution).

\begin{figure}[ht]
\includegraphics[width=0.45\textwidth]{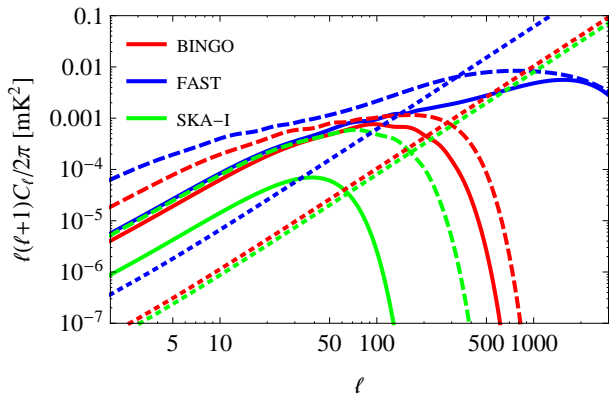}
\caption{\label{fig.cl_nl} The predicted angular power spectra of the lowest frequency bin (solid line), the highest frequency bin (dashed line) listed in Table~\ref{tab.survey_parameter} and the noise levels (thick dotted line) for different surveys. To be specific, the frequency windows used in this plot are $[960,980]$ MHz and $[1240,1260]$ MHz for BINGO, $[1050,1070]$ MHz and $[1330,1350]$ MHz for FAST and $[350,370]$ HMz and $[1030,1050]$ MHz for SKA-I.}
\end{figure}

The finite beam size in observation limits the resolution at small angles. To take into account the effect of beam size, the theoretical angular power spectrum needs to be modulated accordingly,
\begin{equation}
C_{\ell}^{\mathrm{obs}} = C_{\ell}^{\mathrm{th}} e^{-\ell^2 \sigma^2},
\label{eq.Cl_obs}
\end{equation}
where $C_{\ell}^{\mathrm{th}}$ is the theoretical angular power spectrum given by Eq.~(\ref{eq.Cl_th}), $C_{\ell}^{\mathrm{obs}}$ is the predicted 21-cm signal and $\sigma = \theta_{\mathrm{F}}/\sqrt{8\ln 2}$. $\theta_{\mathrm{F}}$ is the full width at half maximum of the instruments
\begin{eqnarray}
\theta_{\rm F}=1.2\left(\frac{\lambda}{D_{\rm dish}}\right),
\end{eqnarray}
where $\lambda=c/\nu$ is the wavelength for the corresponding frequency channel, and $D_{\rm dish}$ is the angular diameter of the dish (Table~\ref{tab.survey_parameter}). In the following, we omit the superscript and use $C_{\ell}$ to refer to $C_{\ell}^{\mathrm{obs}}$. Figure~\ref{fig.cl_nl} shows the signal and noise levels in different surveys assuming the parameters listed in Table~\ref{tab.survey_parameter}. The dotted lines represent the amplitudes of thermal noise $\ell(\ell+1)N_{\ell}$. Note that $N_{\ell}$ is independent of angular scales, hence the lines scale as $\ell(\ell+1)$. The solid lines correspond to the signal of lowest frequency (highest redshift) bins and the dashed lines to the highest frequency (lowest redshift) bins in the frequency range of the respective surveys. The curves are calculated for a model with $w=-1.1$ and zero interaction between dark sectors, which corresponds to the fiducial model in phantom case ($w<-1$) in the forecast below. Although we choose the phantom case here for that the interaction form is less restricted, the results are similar when $w>-1$. The bandwidth of each frequency bin is $\Delta\nu = 20$ MHz. Thus, for example, the red solid line shows the signal of frequency window $[960,980]$ MHz and the red dashed lines show the signal of $[1240, 1260]$ MHz in BINGO. For small $\ell$, the signals trace the brightness temperature angular power spectrum as shown in Fig.~\ref{fig.cl}. Some wiggles can be seen at $\ell$ of a few tens to a few hundreds depending on working frequency (except for the low frequency end of SKA-I for that is beyond its resolution), which corresponds to the BAO in matter power spectrum. With the increase of $\ell$, suppression due to finite beam size becomes important and the signal drops quickly. We can see that the signal of SKA-I is the lowest among the surveys, because its working frequency is the lowest and the signal comes from the highest redshift. Meanwhile, SKA-I also has the lowest noise level, due to the large number of antennae. The worst noise level is found in FAST, but the predicted signal is the strongest, too, because of its focus on low redshifts. However, FAST is able to achieve the highest resolution given its large illuminated aperture, while SKA-I can only detect signals on large angular scales when working in single-dish mode.

\section{Forecast \label{sec.forecast}}

To estimate the potential ability of HI IM in constraining the interaction between DM and DE, we perform a forecast for the future surveys introduced in the previous section using a Fisher matrix analysis. If a model likelihood in the parameter space is a Gaussian distribution, the Fisher matrix equals to the inverse of the covariance matrix of the distribution, i.e. Cramer-Rao inequality~\cite{Tegmark1997}. Even if it is not Gaussian, the Fisher matrix is still a good tool in estimating the real distribution around the mean value, since it can always be well approximated with a Gaussian distribution.

The Fisher matrix element for a HI IM experiment with respect to model parameters $p_{\alpha}$ and $p_{\beta}$, reads
\begin{equation}
F_{\alpha\beta} = f_{\mathrm{sky}} \sum^{\ell_{\mathrm{max}}}_{\ell_{\mathrm{min}}} \bigg( \frac{2\ell +1}{2} \bigg) \mathrm{Tr}[\Gamma_{\ell,\alpha} (\Gamma_{\ell})^{-1} \Gamma_{\ell,\beta} (\Gamma_{\ell})^{-1}],
\label{eq.fisher}
\end{equation}
in which we treat the spherical harmonic decomposition coefficients $a_{lm}$ as the (Gaussian) random variables whose mean value does not depend on cosmological parameters. $f_{\mathrm{sky}}$ is the fractional sky coverage. $\Gamma_{\ell}$ is a matrix composed of the cross and autocorrelation angular power spectra between the frequency windows. $\Gamma^{ij}_{\ell} = C^{ij}_{\ell} + \delta^{ij}N_{\ell}$ is the observed angular power spectrum between window $i$ and $j$, in which $i,j$ run through the frequency windows. The diagonal elements account for the contributions of the autocorrelations of each frequency bin to the Fisher matrix, while the off diagonal elements encompass the contributions of the cross correlations between the frequency bins. The angular power spectrum $C_{\ell}^{ij}$ are computed using Eq.~(\ref{eq.Cl_th}) then modulated according to Eq.~(\ref{eq.Cl_obs}). $\Delta_{T_{\mathrm{b}},\ell}$ in the different frequency (redshift) bins for the same $\ell$ correspond to different comoving scales, hence their cross correlations $C_{\ell}^{ij}$ ($i \neq j$) are much smaller than autocorrelations. Thus the off diagonal elements of $\Gamma_{\ell}$ are usually much smaller than its diagonal elements. Nevertheless we compute all the elements of $\Gamma_{\ell}$ and then use Eq.~(\ref{eq.fisher}) to calculate the Fisher matrix. Following Refs. \cite{Santos2015, Bull2015}, it is assumed that the noise between different windows is uncorrelated which means that the noise only contributes to the diagonal elements of $\Gamma_{\ell}$. ${\Gamma}_{\ell,\alpha}$ (${\Gamma}_{\ell,\beta}$) denotes the partial derivative of $\Gamma_{\ell}$ with respect to the parameter $p_{\alpha}$ ($p_{\beta}$). The noise should not depend on the model parameters, so that $\Gamma_{\ell,\alpha} = C_{\ell,\alpha}$. We set $\ell_{\mathrm{min}} = 2$ as in a CMB experiment. $\ell_{\mathrm{max}}$ depends on the resolution of the instruments. With the increase of $\ell$, the signal to noise ratio decreases as well as the information gain by increasing $\ell_{\mathrm{max}}$ and $F_{\alpha\beta}$ approaches a limit. We set $\ell_{\mathrm{max}}= 1000$ for BINGO and SKA-I and $3000$ for FAST such that the Fisher matrix converges. Although the largest $\ell$'s already lay in the nonlinear region, we ignore nonlinear effects in the analysis. As can be seen in Fig.~\ref{fig.cl_nl}, the signal drops below noise quickly for $\ell$ equals a few hundred. The impact of the nonlinear effect is negligible.

\begin{figure*}[ht]
\subfloat[]{
    \includegraphics[width=0.33\textwidth]{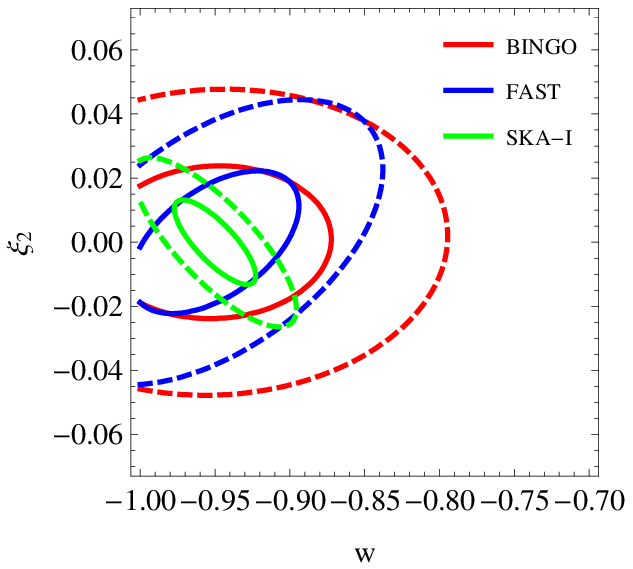}
}
\subfloat[]{
    \includegraphics[width=0.32\textwidth]{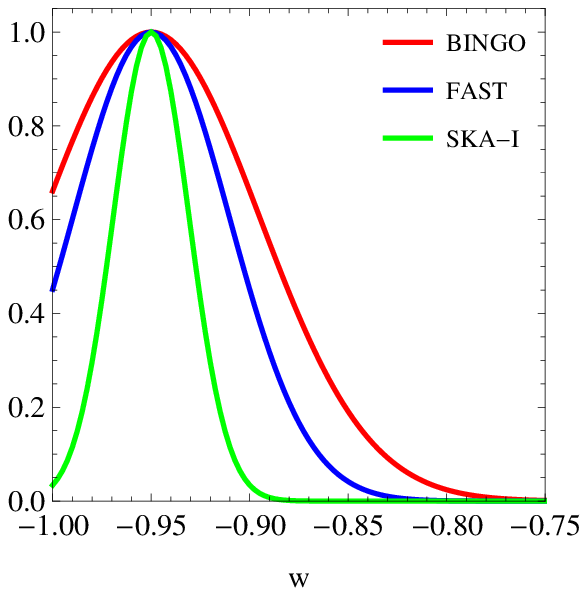}
}
\subfloat[]{
    \includegraphics[width=0.31\textwidth]{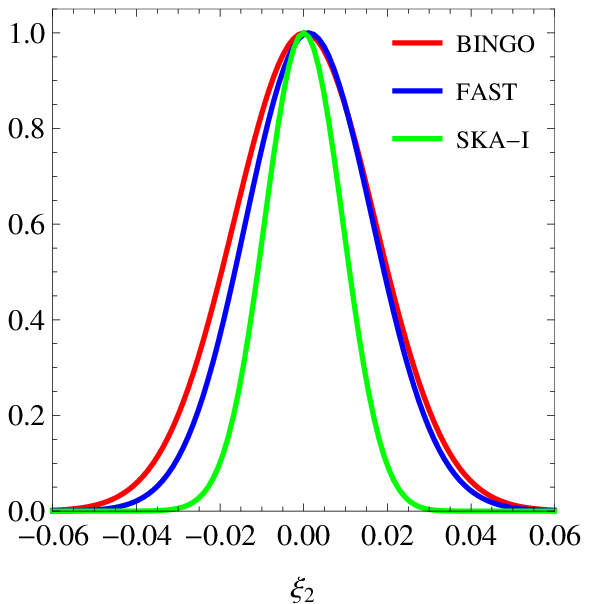}
}
\caption{\label{fig.quint_like} The two dimensional likelihood contours and one dimensional likelihood distributions of $w$ and $\xi_2$ when $w>-1$. In (a), solid and dashed represent 68\% and 95\% C.L. regions respectively. }
\end{figure*}

\begin{figure*}[ht]
\subfloat[]{
    \includegraphics[width=0.31\textwidth]{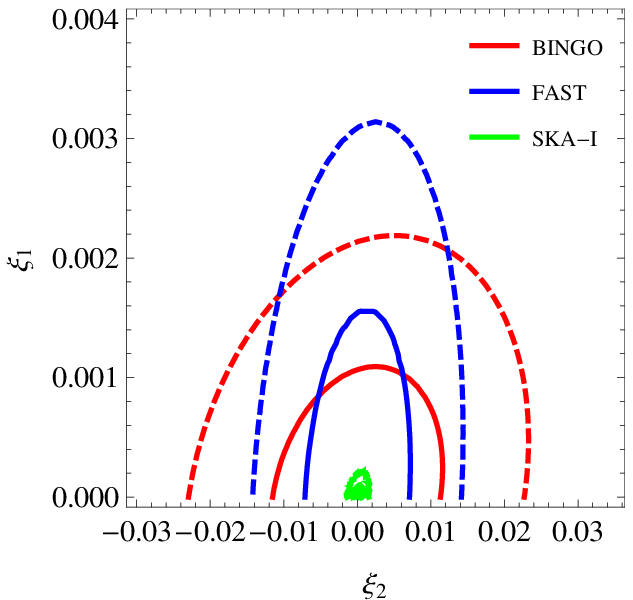}
}
\subfloat[]{
    \includegraphics[width=0.33\textwidth]{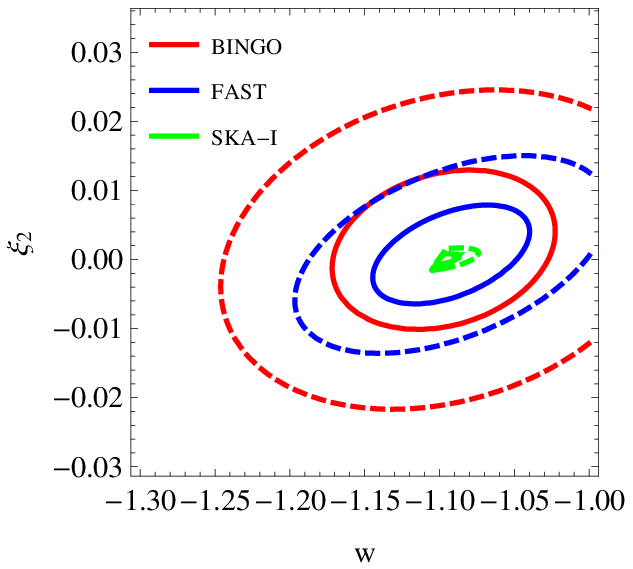}
}
\subfloat[]{
    \includegraphics[width=0.33\textwidth]{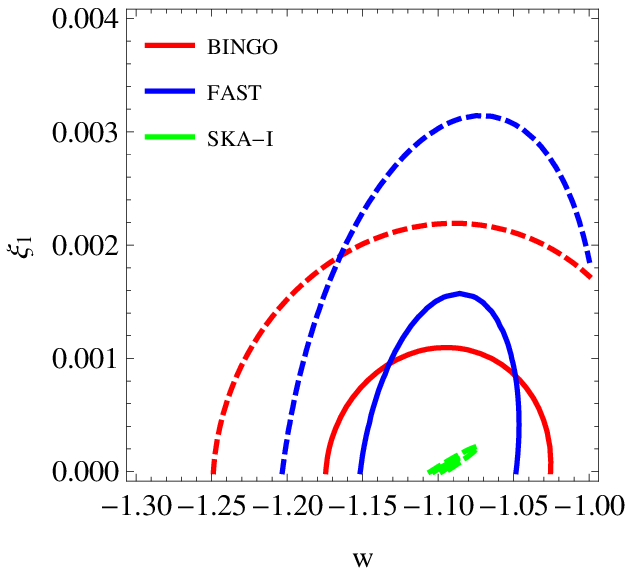}
}
\caption{\label{fig.phant_like2D} The two dimensional likelihood contours among $w$, $\xi_1$ and $\xi_2$ when $w<-1$. Solid and dashed lines represent 68\% and 95\% C.L. regions respectively. In each plot, the likelihood is marginalized over the third parameter.}
\end{figure*}

\begin{figure*}[ht]
\subfloat[]{
    \includegraphics[width=0.3\textwidth]{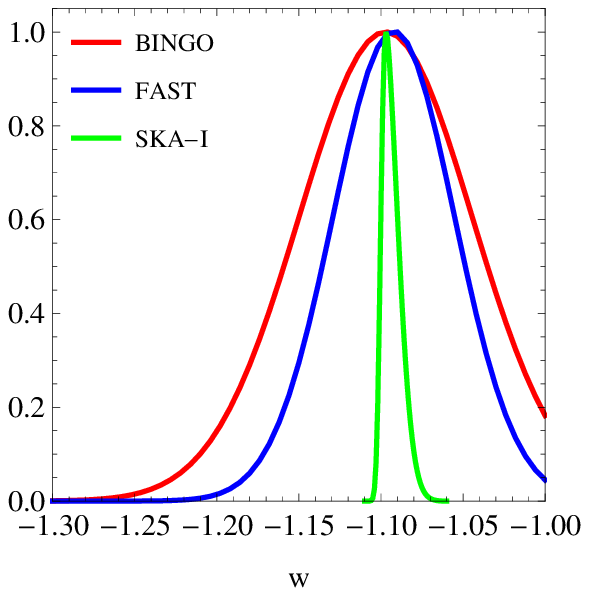}
}
\subfloat[]{
    \includegraphics[width=0.3\textwidth]{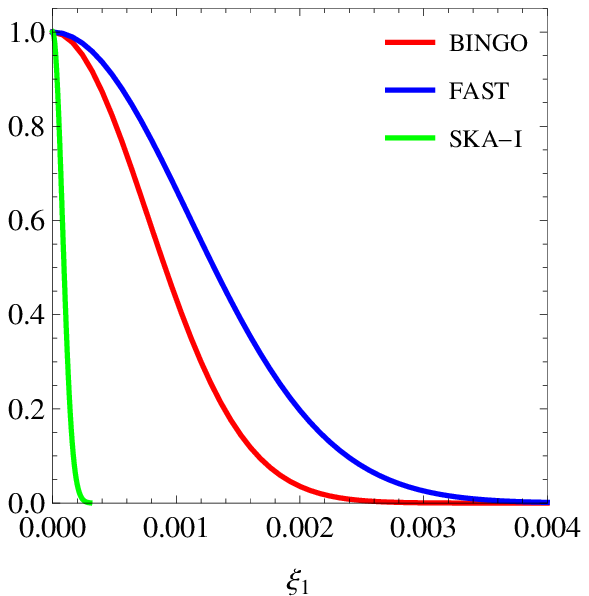}
}
\subfloat[]{
    \includegraphics[width=0.285\textwidth]{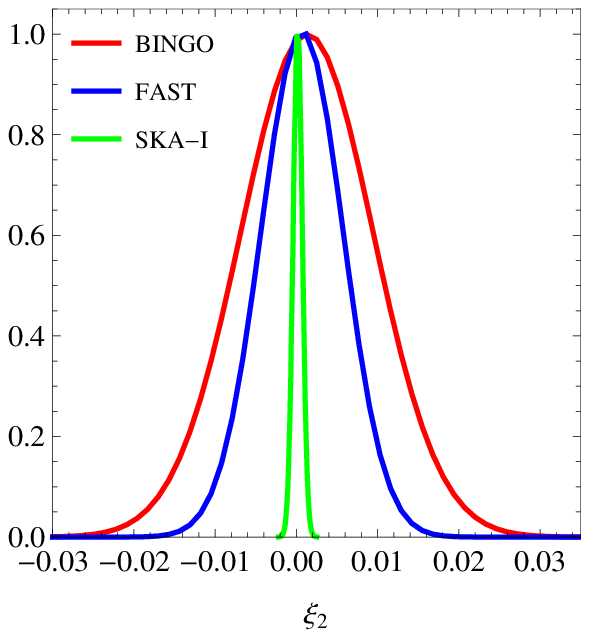}
}
\caption{\label{fig.phant_like1D} The one dimensional likelihood distributions of $w$, $\xi_1$ and $\xi_2$ when $w<-1$.}
\end{figure*}

Figures~\ref{fig.quint_like}--\ref{fig.phant_like1D} show the forecasted two dimensional errors for 68\% and 95\% area as well as marginalized one dimensional probability distributions. The forecasted variances are listed in Table~\ref{tab.constraint}. When $w>-1$, we fix $\xi_1 = 0$ which means that the energy transfer could not be proportional to DM density. When $w<-1$, we vary $\xi_{1,2}$ and $w$ simultaneously. We assume the mean values of the parameters as $w_0=-0.95$ when $w>-1$; $w_0=-1.1$ when $w<-1$ and $\xi_{10}=\xi_{20}=0$. In addition to $w$ and $\xi$'s, we also vary the parameters \{$h$, $\Omega_{\mathrm{b}}h^2$, $\Omega_{\mathrm{c}}h^2$, $n_{\mathrm{s}}$, $10^9A_{\mathrm{s}}(\Omega_{\mathrm{HI}}b_{\mathrm{HI}})^2$\}, where $b_{\mathrm{HI}}$ is the HI bias. The impact of $A_{\mathrm{s}}$, $\Omega_{\mathrm{HI}}$ and $b_{\mathrm{HI}}$ is similar: they modulate the overall amplitude of the 21-cm angular power spectrum, so that we combine them in a single parameter. Since we only consider the redshift after reionization, the power spectrum is insensitive to the optical depth $\tau$; thus we do not include it in our analysis. We set the mean values: $\Omega_{\mathrm{HI}}=0.62\times 10^{-3}$, $b_{\mathrm{HI}}=1$ and the others are set according to the {\it Planck} best fit as listed in the end of Sec.~\ref{sec.intro}. The two dimensional and one dimensional distributions in this section are always marginalized over other parameters. Note that due to the bounds on the parameters, $\xi_1$ must be non-negative and $w$ be either greater or less than $-1$; we cannot observe the whole multivariate Gaussian distribution constructed from the Fisher matrix within the parameter space considered. As a result the marginalized likelihood distributions may not be Gaussian and the inferred mean values can be offset from the mean value we set.

We can see that, for both cases when $w>-1$ and $w<-1$, SKA-I provides the tightest constraints on $w$ and $\xi$'s, though its signal to noise is the lowest, especially on small scales, among the surveys. This demonstrates the advantages of surveys probing a large redshift range in testing interacting dark energy models. When $w>-1$, SKA-I constrains $w$ and $\xi$'s about twice as tightly as BINGO, while its constraints are one magnitude better when $w<-1$. Under our assumption that the coupling coefficients are constant, the interaction proportional to DM density is forbidden when $w>-1$. On the other hand, when $w<-1$ it dominates over the interaction proportional to DE density in matter era, where BINGO or FAST cannot detect but overlaps with the redshift range covered by SKA-I. Therefore SKA-I is uniquely suitable in detecting such interaction, leading to the outstanding performance in the case $w<-1$. Comparing results of BINGO and FAST, FAST usually gives marginally tighter constraints, thanks to its high sensitivities, except for $\xi_1$ when $w<-1$. Again, this is a consequence of the fact that BINGO probes higher redshift than FAST where the interaction proportional to DM density is more prominent.

Apart from the magnitude of the variance, orientations of the distributions also vary with surveys, because of the different redshift ranges that these survey observe. Consider the background expansion of the Universe, or Friedmann equation, in IDE cosmology. Substituting Eq.~(\ref{eq.Q}) into Eq.~(\ref{eq.rhoc}) and Eq.~(\ref{eq.rhod}), the evolution of the homogeneous DM and DE in the presence of the interaction is effectively equivalent to two noninteracting fluids with EoS $w_{\mathrm{c}} = -\xi_1-\xi_2/r$ and $w_{\mathrm{d}} = w+\xi_1r-\xi_2$, where $r \equiv \rho_{\mathrm{c}}/\rho_{\mathrm{d}}$. The influence of the coupling coefficients and DE EoS is degenerated and the degeneracy is time dependent through $r$. Therefore the orientation of the probability distribution in $(w,\xi_1,\xi_2)$ space in one survey appears to be different than the other survey due to the different frequency ranges.

\begin{table*}[tb!]
\caption{\label{tab.constraint} The forecasted errors of the model parameters.}
\begin{tabular}{p{15pt}p{110pt}p{110pt}p{110pt}}
    \toprule
    & BINGO & FAST & SKA-I \\
    \cline{2-4}
    & \multicolumn{3}{c}{$w>-1$}  \\
    \cline{2-4}
    $\sigma_{w}$     & $4.03 \times 10^{-2}$ & $3.24 \times 10^{-2}$ & $1.83 \times 10^{-2}$ \\
    $\sigma_{\xi_1}$ & N/A                   & N/A                   & N/A \\
    $\sigma_{\xi_2}$ & $1.60 \times 10^{-2}$ & $1.44 \times 10^{-2}$ & $8.86 \times 10^{-3}$ \\
    \hline
    & \multicolumn{3}{c}{$w<-1$} \\
    \cline{2-4}
    $\sigma_{w}$     & $4.86 \times 10^{-2}$ & $3.51 \times 10^{-2}$ & $5.13 \times 10^{-3}$ \\
    $\sigma_{\xi_1}$ & $7.23 \times 10^{-4}$ & $1.05 \times 10^{-3}$ & $7.53 \times 10^{-5}$ \\
    $\sigma_{\xi_2}$ & $7.77 \times 10^{-3}$ & $4.80 \times 10^{-3}$ & $5.52 \times 10^{-4}$ \\
    \lasthline
\end{tabular}
\end{table*}

\begin{table*}[tb!]
\caption{\label{tab.Planck_fit} Fitting results in~\cite{Costa2017} and comparison to the forecast results. The quoted error is for $68\%$ confidence level. }
\begin{tabular}{p{15pt}p{110pt}p{110pt}p{110pt}p{110pt}}
    \toprule
    & Model I & Model II & Model III & Model IV \\
    & ($w>-1$, $\xi_{1}=0$, $\xi_{2}<0$) & ($w<-1$, $\xi_{1}=0$, $\xi_{2}>0$) & ($w<-1$, $\xi_{1}>0$, $\xi_{2}=0$) & ($w>-1$, $\xi_{1}=\xi_{2}>0$) \\ \cline{2-5}
    & \multicolumn{4}{c}{{\it Planck} CMB only}  \\
    \cline{2-5}
    $w$     & $-9.031^{+0.23}_{-0.959} \times 10^{-1}$ & $-1.55^{+0.235}_{-0.358}$ & $-1.702^{+0.298}_{-0.364}$ & $-1.691^{+0.318}_{-0.359}$ \\
    $\xi_1$ & N/A & N/A & $1.458^{+0.37}_{-1.46} \times 10^{-3}$ & $1.416^{+0.37}_{-1.42} \times 10^{-3}$ \\
    $\xi_2$ & $-1.297^{+1.30}_{-0.448} \times 10^{-1}$ & $3.88^{+1.16}_{-3.88} \times 10{-2}$ & N/A & $1.416^{+0.37}_{-1.42} \times 10^{-3}$ \\
    \hline
    & \multicolumn{4}{c}{{\it Planck} CMB+BAO+SNIa+H0+growth rate} \\
    \cline{2-5}
    $w$     & $-9.541^{+0.188}_{-0.372} \times 10^{-1}$ & $-1.035^{+0.0341}_{-0.00835}$ & $-1.069^{+0.0268}_{-0.0152}$ & $-1.07^{+0.0284}_{-0.0163}$ \\
    $\xi_1$ & N/A & N/A & $6.628^{+2.41}_{-5.92} \times 10^{-4}$ & $7.587^{+3.35}_{-6.02} \times 10^{-4}$ \\
    $\xi_2$ & $-1.815^{+1.82}_{-0.328} \times 10^{-3}$ & $2.047^{+0.656}_{-0.667} \times 10^{-2}$ & N/A & $7.587^{+3.35}_{-6.02} \times 10^{-4}$ \\
    \lasthline
\end{tabular}
\end{table*}

In Table~\ref{tab.Planck_fit} we listed the fitting results of IDE model reported in \cite{Costa2017} using two data sets: CMB data only from {\it Planck} 2015~\cite{PlanckCollaboration2015a} and a combined data set of {\it Planck} 2015, BAO measurements from 6dF Galaxy Survey and Sloan Digital Sky Survey \cite{Anderson2013, Ross2014, Beutler2011}, the Joint Lightcurve Analysis SNIa data \cite{Betoule2014}, a recent local Hubble parameter measurement \cite{Efstathiou2013} and a compilation of large scale structure growth rate measurements from redshift space distortion (RSD) and peculiar velocity observations. In Ref.~\cite{Costa2017}, four subclasses of the IDE model were investigated: Model I where $w>-1$, $\xi_1=0$ and $\xi_2<0$, Model II where $w<-1$, $\xi_1=0$ and $\xi_2>0$, Model III where $w<-1$, $\xi_1>0$ and $\xi_2=0$ and Model IV where $w>-1$, $\xi_1=\xi_2>0$. As we can see, our forecasted constraints of all three surveys by using HI IM are comparable with or tighter than those from {\it Planck}, which is the most precise CMB experiment to date. Furthermore, HI IM of SKA-I still slightly outperforms the combined data sets, which proves it to be a promising probe of the interaction between dark sectors. But we must point out that our analysis is an ideal one in the sense that we ignore the contamination of foreground hence the noise level is underestimated, although foreground removal techniques may minimize the residual noise. However, as combining {\it Planck} data with SN, BAO and RSD data improves the constraints on dark energy models because the degeneracies between the cosmological parameters, $w$ and $\xi$'s for example, are broken by fitting to low and high redshift observations simultaneously \cite{He2010a}, we expect that combining HI intensity with CMB measurements will break the degeneracies as well and provide even better results. In addition, the working frequency ranges of BINGO, FAST and SKA-I are complementary to each other so the combined constraints will be stronger and more reliable.

\begin{figure*}[ht]
\subfloat[]{
    \includegraphics[width=0.47\textwidth]{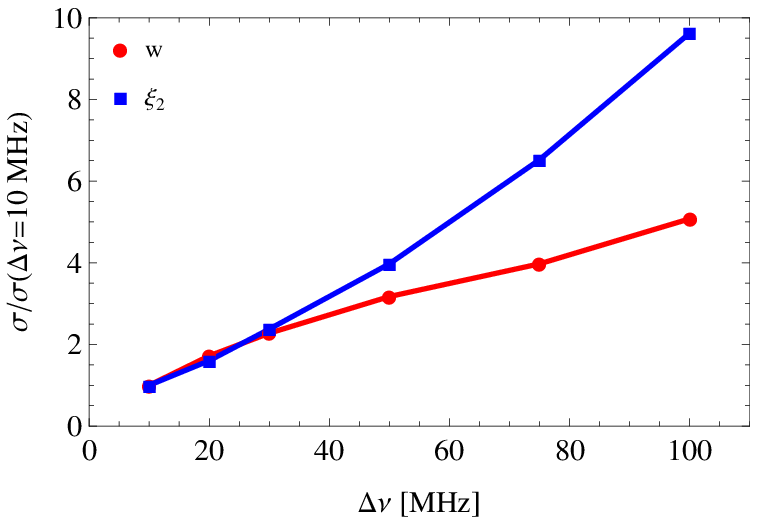}
}
\subfloat[]{
    \includegraphics[width=0.45\textwidth]{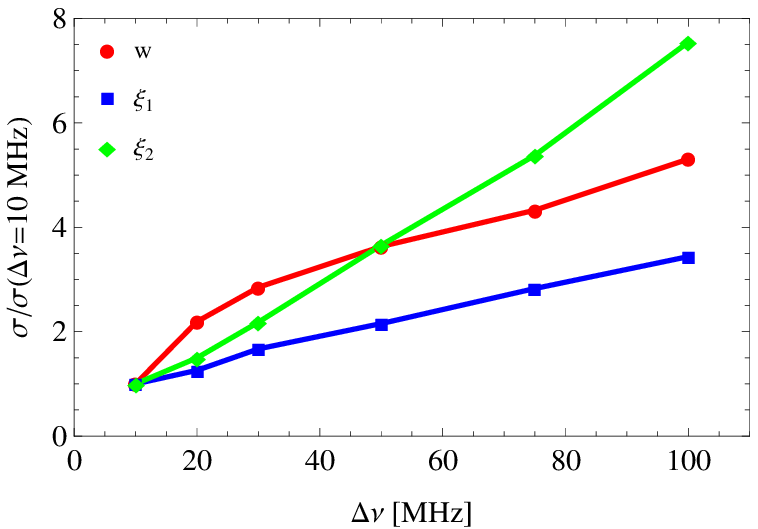}
}
\caption{\label{fig.delta_nu_var} The forecasted errors of $w$ and $\xi_{1,2}$ assuming different window width $\Delta\nu$. In (a) $w>-1$ and (b) $w<-1$. The errors are normalized to the case $\Delta\nu=10$\,MHz, for which $\sigma_{w} = 2.34 \times 10^{-2}$ and $\sigma_{\xi_2} = 1.00 \times 10^{-2}$ when $w > -1$, and $\sigma_{w} = 2.22 \times 10^{-2}$, $\sigma_{\xi_1} = 5.74 \times 10^{-4}$ and $\sigma_{\xi_2} = 5.19 \times 10^{-3}$ when $w < -1$. }
\end{figure*}

Given Eq.~(\ref{eq.noise}), the noise level is proportional to the inverse of frequency window width $\Delta\nu$. A wide window would suppress thermal noise. Meanwhile, a narrower window can effectively boost up the RSD contribution to the 21-cm power spectrum as well as the observed signal. To investigate the influence of window width on the constraints, we compare six configurations based on BINGO in which $\Delta\nu$ are set to $10$, $20$, $30$, $50$, $75$ and $100$ MHz. The other experimental parameters are the same as in Table~\ref{tab.survey_parameter}. The results are plotted in Fig.~\ref{fig.delta_nu_var}. It is clear that a narrower window gives rise to better constraints. This implies that, though the thermal noise is enhanced while decreasing window width, the signal to noise is increased. A narrower window also leads to more detailed slices in the observed redshift range, which may help in tightening the constraints.

\begin{figure*}[ht]
\subfloat[]{
    \includegraphics[width=0.45\textwidth]{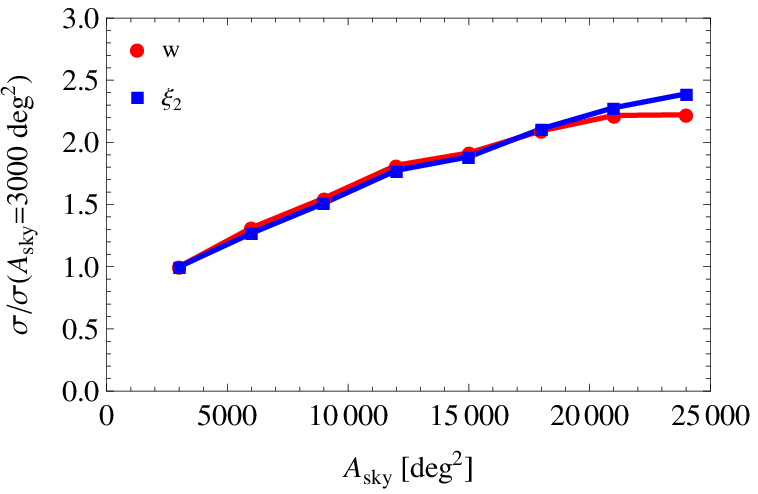}
}
\subfloat[]{
    \includegraphics[width=0.45\textwidth]{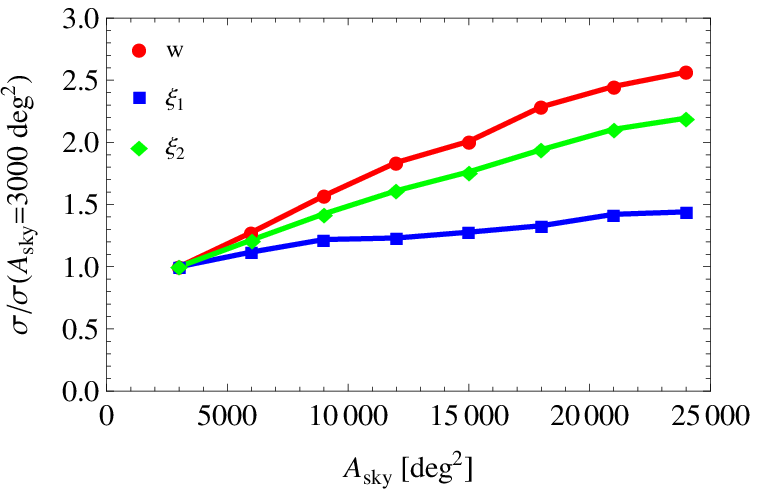}
}
\caption{\label{fig.sky_var} The forecasted errors of $w$ and $\xi_{1,2}$ assuming different sky coverage $A_{\mathrm{sky}}$. In (a) $w>-1$ and (b) $w<-1$. The errors are normalized to the case $A_{\mathrm{sky}}=3000\,\mathrm{deg}^2$, for which $\sigma_{w} = 1.46 \times 10^{-2}$ and $\sigma_{\xi_2} = 6.03 \times 10^{-3}$ when $w > -1$, and $\sigma_{w} = 1.37 \times 10^{-2}$, $\sigma_{\xi_1} = 7.26 \times 10^{-4}$ and $\sigma_{\xi_2} = 2.19 \times 10^{-3}$ when $w < -1$. }
\end{figure*}

Another parameter similar to frequency window width is sky coverage $A_{\mathrm{sky}}$, assuming a given total observation time. The larger the sky coverage is, the smaller the period that each pixel is sampled, and therefore the higher the pixel noise is. In contrast, the larger the sky coverage is, the larger the volume of the Universe that the survey samples, and the smaller the cosmic variance is. Thus there is a competition between pixel noise and cosmic variance for different values of $A_{\rm sky}$. Taking FAST as an example, we test the influence of $A_{\mathrm{sky}}$ on the constraints in Fig.~\ref{fig.sky_var}, where we show the forecasted relative variances of eight configurations with sky coverage ranging from $3000$ to $24000~\mathrm{deg}^2$. The results show that smaller $A_{\mathrm{sky}}$ provide better constraints, i.e. better precision in the angular power spectrum measurements. From Fig.~\ref{fig.cl}, we can see that the deviation of $C_{\ell}$ in IDE models from $\Lambda$CDM rises at large $\ell$, where cosmic variance is subdominant compared to instrumental noise. In terms of probing interaction between dark sectors, a precise survey focusing on a smaller sky area is more efficient than a larger sky-area survey with higher pixel noises.

\section{Conclusion \label{sec.conclusion}}

In this work we investigate the potential constraints on the interaction between DM and DE by using HI intensity mapping observations. We show that the 21-cm cross and autocorrelation angular power spectra show features in the presence of such interaction. Considering three ongoing surveys, BINGO, FAST and SKA-I, we find that their constraints on the interaction can be comparable or better than current results using CMB, BAO, SNIa, local Hubble parameter and growth rate data in an optimistic situation where the foreground is completely removed. Thus HI intensity mapping is a promising tool, potentially more efficient than any single measurement available, in probing dark sector interactions. The constraints of certain models are affected by the survey configurations, including the splitting of frequency bins and sky coverage. Since the difference between standard $\Lambda$CDM predictions and interacting models is mostly shown on small scales, lowering the pixel noises can more effectively strengthen the constraints. Therefore, a smaller sky coverage of the survey can be more effective in tightening up the constraints.

\begin{acknowledgments}
We thank Yi-Chao Li for helpful discussions. This work is based on the research supported by the South African Research Chairs Initiative of the Department of Science and Technology and the National Research Foundation (NRF) of South Africa (A. W. and X. X.) as well as the Competitive Programme for Rated Researchers (Grant No. 91552) (A. W.). Y.-Z. M. acknowledges the support of the NRF of South Africa with Grant No.105925. A. W. thanks the Centre for Computational Astrophysics and the Institute for Advanced Study for hosting her while the work was in progress. Any opinion, finding and conclusion or recommendation expressed in this material is that of the authors and the NRF of South Africa does not accept any liability in this regard.
\end{acknowledgments}

\bibliographystyle{apsrev4-1}
\bibliography{HIintensity}

\end{document}